\documentclass[twocolumn,showpacs,superscriptaddress,floatfix]{revtex4}
\usepackage{graphicx}
\usepackage{bm}

\begin{document}

\title{A multiband envelope function model for
quantum transport in a tunneling diode}

\author{Omar Morandi}
\email{omar.morandi@unifi.it} 
\affiliation{ Dipartimento di Elettronica e Telecomunicazioni, 
Universit\`a di Firenze, \\
Via S. Marta 3, 50139 Firenze, Italy}
\author{Michele Modugno}
\affiliation{LENS, Dipartimento di Fisica and INFM, Universit\`a di Firenze, \\
Via Nello Carrara 1, 50019 Sesto Fiorentino, Italy}

\date{\today}

\begin{abstract}
We present a simple model for electron transport in semiconductor
devices that exhibit tunneling between the conduction and valence
bands.  The model is derived within the usual Bloch-Wannier
formalism by a $k$-expansion, and is formulated in terms of a set
of coupled equations for the electron envelope functions.  Its
connection with other models present in literature is discussed.
As an application we consider the case of a Resonant Interband
Tunneling Diode, demonstrating the ability of the model to
reproduce the expected behaviour of the current as a function of
the applied voltage.
\end{abstract}

\maketitle

\section{Introduction}
\label{sec:intro}

In recent years there has been a growing interest for
semiconductor devices characterized by tunneling effects between
different bands, as the Resonant Interband Tunneling Diode (RITD)
\cite{Genoe}. This kind of diode belongs to the class of
heterostructures that show a negative differential resistance in a
certain range of the applied voltage, like the widely employed RTD
(Resonant Tunneling Diode) \cite{KLU:89,Buot}. However,
differently from the latter where the electronic current flows
within a single band, the remarkable feature of a RITD is the
possibility to achieve a sharp coupling between ``conduction" and
``valence" states, allowing an interband current which becomes the
main transport phenomena in the resonant region.

The description of electron transport in such quantum devices
hence requires multiband models capable to account for tunneling
mechanisms between different bands induced by the heterostructure
design and the applied external bias.

In the literature, different methods are currently employed for
characterizing the band structures and the electronic or optical
properties of these heterostructures, such as envelope functions
methods based on the effective mass theory
\cite{Chao,McGill2,McGill3}, tight-binding \cite{Boyk,Fens1} and
pseudopotential \cite{Bach} methods. In addition, various
mathematical tools are employed to exploit the multiband quantum
dynamics underlying the previous models: the Schr\"odinger-like
models \cite{swexu}, the nonequilibrium Green's function
\cite{Lake,Lake2}, the Wigner function approach
\cite{kanettsp,santafe2001,modena2003}, and recently the
hydrodynamics multiband formalisms \cite{alifroman,barletti}.

All of these methods rely on some common approximations to account
for the effects of a non-uniform band profile on the electron
dynamics. In particular, in the usual ``${\bf k} \cdot {\bf P}$''
approach \cite{callaway}, one starts by defining the Hamiltonian
matrix of the bulk (in ${\bm k}$-space), and then allows the
physical parameters (typically the band eigenvalues or the
Luttinger-Kohn parameters \cite{luttinger}) to have some ${\bm
x}$-dependence in order to describe the position-dependent
properties of the heterostructure. In this approach care must be
taken to preserve the self-jointness of Hamiltonian matrix, so
appropriate quantization rules are needed \cite{Chao_Chuang}. In
this way, the electrical fields arising from the band edge offset
among different layers are not included from the beginning in the
derivation of the model, but appear only at the macroscopic level
(i.e. at the level of envelope functions). Indeed, in the previous
approximation technique, the ${\bm x}$-dependence of the
unperturbed Hamiltonian matrix elements, generates a ``mean
effective electric field" acting on the envelope functions, which
is not present at the microscopic level.

A different approach has been proposed in \cite{Foreman} were a
local ``modified Wannier basis" is chosen to include the
inhomogeneity directly into the basis elements. Unfortunately, in
this case the equations of motion of the envelope functions depend
on the change of the Bloch functions across the interfaces (that
are implicitly neglected in the previous procedure) and such an
evaluation can result in a very difficult task.

In this paper we introduce a different strategy, describing the
band edge offsets by means of external potentials applied to the
bulk structure. This allows us to treat on the same footing both
the electrostatic potential generated by the charge distribution
in the device and the heterostructure design of band edges, in
order to highlight the role played by the heterostructure
potential in the interband tunneling process.

Within this framework we derive a hierarchy of multiband models
obtained by means of a $k$-expansion, were the momentum $k$ plays
the role of asymptotic parameter as in the usual ``${\bf k} \cdot
{\bf P}$'' approach. The starting point is the single electron
Bloch representation, that here we consider for simplicity for the
case of non degenerate bands and constant band gaps. Then, after
the $k$-expansion, the electron wavefunction is projected on the
Wannier basis, yielding a set of coupled Schr\"odinger equations
for the electron envelope functions in definite energy bands.

These equations share some similarities with those of the well
known Kane \cite{kane} and Luttinger-Kohn (LK) \cite{luttinger}
models. However, a key difference is the choice of the basis
elements. Indeed, since in a uniform crystal the Wannier functions
of a given energy band are related to the Bloch functions of the
same band by a unitary transformation, this allows us to give a
simple physical insight to the envelope functions. Differently,
since the Kane model arises form a unitary transformation of only
the periodic part of the Bloch functions, the generic element of
the Kane basis is nondiagonal in the Bloch band index $n$, and
envelope functions related to different ``band" indices turn out
to be coupled even in absence of any applied potential, therefore
lacking of a direct physical interpretation.

The LK model instead is a multiband effective mass model obtained
from the latter by an additional quasi-unitary transformation that
removes the spurious interband coupling to first order in $k$.
However, since the LK approach is devoted to describe intraband
effects, the coupling due to the external field is generally
neglected.

As an application of the present approach we consider the case of
a two-band RITD, showing that the model is able to reproduce the
expected behaviour of the current as a function of the applied
voltage.

The paper is organized as follows. In the next section we discuss
the derivation of the model and the approximations employed; then
in Sec. \ref{sec:comparison} we analyze differences and analogies
with the Kane \cite{kane} and Luttinger-Kohn \cite{luttinger}
models.  Finally, in Sec. \ref{sec:example} we work out explicitly
the case of a RITD, investigating its current-voltage
characteristic curve.

\section{Derivation of the model}
\label{sec:model}

Let us consider an electron of mass $m$ immersed in a crystal
lattice described by the periodic potential $V_L$, in the presence
of an additional external potential $U$ that will be treated as a
perturbation.  The evolution of the electron wavefunction
$\Psi(\textbf{x},t)$ is given by the solution of the Schr\"odinger
equation
\begin{equation}
i\hbar\partial_t\Psi(\textbf{x},t)=\left[
  -\frac{\hbar^2}{2m}\nabla^2+V_L(\textbf{x})+U(\textbf{x})\right]
\Psi(\textbf{x},t)\,. \label{eq:schrod}
\end{equation}

The eigenfunction of the unperturbed Hamiltonian
$H_0=-(\hbar^2/2m)\nabla^2+V_L$ are Bloch functions
$\psi_n(\textbf{k},\textbf{x})$ (see e.g. \cite{callaway})
\begin{equation}\label{eq:Bloch_eigen}
H_0\psi_n(\textbf{k},\textbf{x})=E_n(\textbf{k})\psi_n(\textbf{k},\textbf{x})
\end{equation}
and form a complete set with the orthonormality condition
\begin{equation}
\int_x\psi_n^*(\textbf{k},\textbf{x})\psi_{n'}(\textbf{k}',\textbf{x})=
\delta(\textbf{k}\!-\!\textbf{k}')\delta_{nn'}
\label{eq:orthonorm}
\end{equation}
$n$ being the band index and $\textbf{k}$ the electron
quasimomentum. Eq. (\ref{eq:schrod}) can be transformed in
momentum space by means of standard textbook methods
\cite{ashcroft,wenckebach} that we review here below for
completeness.  According to the Bloch theorem the Bloch functions
$\psi_n(\textbf{k},\textbf{x})$ can be written as
\begin{equation}
\psi_n(\textbf{k},\textbf{x})={\rm
e}^{i\textbf{k}\cdot\textbf{x}}u_n(\textbf{k},\textbf{x}) \equiv
\langle\textbf{x}|n,\textbf{k}\rangle \label{eq:eigen}
\end{equation}
where the functions $u_n(\textbf{k},\textbf{x})$ have the same
periodicity of the lattice potential and are normalized according
to
\begin{equation}
\frac{(2\pi)^3}{\Omega}\int_{cell}u_n^*(\textbf{k},\textbf{x})
u_{n'}(\textbf{k},\textbf{x}) =\delta_{nn'}.
\end{equation}

A generic solution of Eq. (\ref{eq:schrod}) can be expanded as
\begin{equation}
\Psi(\textbf{x},t)=
\sum_n\int_k\varphi_n(\textbf{k},t)\psi_n(\textbf{k},\textbf{x})
\end{equation}
where $\textbf{k}$ runs over the first Brillouin zone; then the
expansion coefficients satisfy the following equation (hereinafter
we omit the time dependence to simplify the notation)
\begin{equation}
i\hbar\partial_t\varphi_n(\textbf{k})=E_n(\textbf{k})\varphi_n(\textbf{k})+
\sum_{n'}\int_{k'}\langle
n,\textbf{k}|U|n',\textbf{k}'\rangle\varphi_{n'}(\textbf{k}')\,.
\label{eq:coeff}
\end{equation}

By exploiting the periodicity of the $u_n(\textbf{k},\textbf{x})$
functions the expectation value of the external potential $U$ can
be rewritten as \cite{callaway}
\begin{eqnarray}
&&\langle n,\textbf{k}|U|n',\textbf{k}'\rangle=\int_x {\rm
e}^{i(\textbf{k}'-\textbf{k})\cdot\textbf{x}}
u_n^*(\textbf{k},\textbf{x})u_{n'}(\textbf{k}',\textbf{x})U(\textbf{x})
\nonumber\\ &&\quad=\sum_l B_l(n,n',\textbf{k},\textbf{k}') \int_x
{\rm e}^{i(\textbf{k}'-\textbf{k}
-\textbf{K}_l)\cdot\textbf{x}}U(x)\nonumber\\
&&\quad=(2\pi)^3\sum_l B_l\tilde{U}(\textbf{k}'-\textbf{k}
-\textbf{K}_l)
\end{eqnarray}
$\textbf{K}_l$ being a reciprocal lattice vector, and $\tilde{U}$
the Fourier transform of $U$.

At this point, following \cite{wenckebach}, we assume the
potential $U$ to be nearly constant over a single lattice cell, so
that only the zero momentum Fourier component give a relevant
contribution
\begin{equation}
\langle n,\textbf{k}|U|n',\textbf{k}'\rangle\simeq(2\pi)^3
B_0\tilde{U}(\textbf{k}'-\textbf{k}) \label{eq:u}
\end{equation}
where $B_0$ can be expressed as
\begin{equation}
B_0=\frac{1}{\Omega}\int_{cell}
u_n^*(\textbf{k},\textbf{x})u_{n'}(\textbf{k}',\textbf{x}) \equiv
\langle u_{n,\textbf{k}}|u_{n',\textbf{k}'}\rangle
\end{equation}
$\Omega$ being the volume of a single cell.

Let us now evaluate explicitly the coefficients $B_0$, by
considering separately the case $n=n'$ and $n\neq n'$. In the
former case it is easy to show from Eq. (\ref{eq:orthonorm}) that
\begin{equation}
B_0(n,n,\textbf{k},\textbf{k}')=1/(2\pi)^3
\end{equation}
with the assumption that both $\textbf{k}$ and $\textbf{k}'$ lie
within the first Brillouin zone so that their difference is not a
reciprocal lattice vector \cite{callaway}.

The case $n\neq n'$ can be carried out by considering the
eigenvalue\ equation for the $u_n(\textbf{k},\textbf{x})$
functions
\begin{equation}
\bar{H}_0(\textbf{k}) |u_{n,\textbf{k}}\rangle = E_n(\textbf{k})
|u_{n,\textbf{k}}\rangle \label{eq:eigenu}
\end{equation}
where we have defined $\bar{H}_0(\textbf{k})$ as
($\hat{\textbf{p}}\equiv-i\hbar{\bm\nabla}$)
\begin{equation}
\bar{H}_0(\textbf{k}) \equiv \frac{1}{2m}(\hat{\textbf{p}}+
\hbar\textbf{k})^2 + V_{L}(\textbf{x})
\end{equation}
Then, by left multiplying Eq. (\ref{eq:eigenu}) by $\langle
u_{n',\textbf{k}'}|$ and using the equivalence
\begin{equation}
\bar{H}_0(\textbf{k}) =\bar{H}_0(\textbf{k}') +
\frac{\hbar^2}{2m}(k^2-k'^2)+
\frac{\hbar}{m}\hat{\textbf{p}}\!\cdot\!(\textbf{k}\!-\!\textbf{k}')
\end{equation}
we get
\begin{eqnarray}
B_0(n,n'\neq n,\textbf{k},\textbf{k}')=
\frac{\hbar}{m}(\textbf{k}\!-\!\textbf{k}')
\frac{\textbf{P}_{nn'}(\textbf{k},\textbf{k}')/(2\pi)^3} {\Delta
E_{nn'}(\textbf{k},\textbf{k}')} \label{eq:b0p}
\end{eqnarray}
with the momentum matrix elements
$P_{nn'}(\textbf{k},\textbf{k}')$ defined by
\begin{equation}
\textbf{P}_{nn'}(\textbf{k},\textbf{k}')
\equiv\frac{(2\pi)^3}{\Omega}\int_{cell}
u_n^*(\textbf{k},\textbf{x})(-i\hbar{\bm\nabla})u_{n'}(\textbf{k}',\textbf{x})
\end{equation}
and
\begin{equation}
\Delta E_{nn'}(\textbf{k},\textbf{k}')\equiv
E_{n}(\textbf{k})-E_{n'}(\textbf{k}')-\frac{\hbar^2}{2m}
\left(k^{2}\!-\!k'^2\right)
\end{equation}

Finally, the equation (\ref{eq:coeff}) for the expansion
coefficients can be rewritten as
\begin{eqnarray}
\label{eq:coeff2} &&
i\hbar\partial_t\varphi_n(\textbf{k})=E_n(\textbf{k})\varphi_n(\textbf{k})+
\int_{k'}\tilde{U}(\textbf{k}\!-\!\textbf{k}')\varphi_n(\textbf{k}') \\
&&\quad+\frac{\hbar}{m} \sum_{n'\neq
n}\int_{k'}\frac{\textbf{P}_{nn'}(\textbf{k},\textbf{k}')}{\Delta
E_{nn'}(\textbf{k},\textbf{k}')} (\textbf{k}\!-\!\textbf{k}')
\tilde{U}(\textbf{k}\!-\!\textbf{k}')\varphi_{n'}(\textbf{k}')
\nonumber\end{eqnarray} where it is easy to identify the single
band dynamics (first line) and the interband coupling (second
line). This equation is so far very general and relies on the only
assumption that the external potential $U$ as no appreciable
variation on the scale of a single lattice cell (see Eq.
(\ref{eq:u})).

Let us now transform back the above equation in coordinate space;
this can be achieved by projection on the Wannier basis
\begin{equation}
\Psi(\textbf{x})=\sum_n\sum_{\textbf{R}_i}
\chi_n(\textbf{R}_i)\phi_n^W(\textbf{x}\!-\!\textbf{R}_i)
\label{eq:wannier}
\end{equation}
where the Wannier basis functions satisfy the orthogonality
relation
\begin{equation}
\int_x\phi_n^{W*}(\textbf{x}\!-\!\textbf{R}_i)\phi_{n'}^W(\textbf{x}\!-\!\textbf{R}_j)
=\delta_{nn'}\delta_{ij} \label{eq:complete_w}
\end{equation}
and can be expressed in terms of Bloch functions as
\begin{equation}
\phi_n^W(\textbf{x}\!-\!\textbf{R}_i)=\sqrt{\frac{\Omega}{(2\pi)^3}}
\int_k\psi_n(\textbf{k},\textbf{x}\!-\!\textbf{R}_i).
\end{equation}
The use of the Wannier basis has two advantages: (i) the
amplitudes $\chi_n(\textbf{R}_i)$, that play the role of envelope
functions on the new basis (see Eq. (\ref{eq:wannier})), can be
obtained from the Bloch coefficients in Eq. (\ref{eq:coeff2}) by a
simple Fourier transform
\begin{equation}\label{eq:def_env_f}
\chi_n(\textbf{R}_i)=\sqrt{\frac{\Omega}{(2\pi)^3}}
\int_k\varphi_n(\textbf{k}){\rm e}^{i\textbf{k}\cdot\textbf{R}_i};
\end{equation}
(ii) they can be interpreted as the actual wave function of an
electron in the $n$th band if one is interested in ``macroscopic''
properties of the system on a scale much larger that the lattice
spacing (that is equivalent to average on a scale of the order of
the lattice cell).  For example, by using the completeness of the
Wannier basis in Eq. (\ref{eq:complete_w}), the density and
current distributions can be expressed as
\begin{eqnarray}
\bar{\rho}_i&\equiv&\langle\rho(\textbf{x})\rangle_{cell-i}\simeq
\sum_{n}|\chi_n(\textbf{R}_i)|^2 \\
\bar{\textbf{J}}_i&\equiv&\langle\textbf{J}(\textbf{x})\rangle_{cell-i}
\simeq\frac{\hbar}{im}\textrm{Im}\sum_{n}[\chi_n^*(\textbf{R}_i)
\bm{\nabla}\chi_n(\textbf{R}_i)]
\end{eqnarray}
Since the functions $\chi_n(\textbf{R}_i)$ are in principle
defined only at the lattice sites, it is convenient to follow the
approach of \cite{adams} and perform the limit to the continuum by
extending the dependence of the $\chi_n(\textbf{R}_i)$ to the
whole space ($\textbf{R}_i\longrightarrow\textbf{x}$). This yields
the following expressions for the cell-averaged charge and current
densities
\begin{eqnarray}
\bar{\rho}(\textbf{x})&\simeq&\sum_{n}|\chi_n(\textbf{x})|^2 \\
\bar{\textbf{J}}(\textbf{x})&\simeq&
\frac{\hbar}{im}\textrm{Im}\sum_{n}[\chi_n^*(\textbf{x})
\bm{\nabla}\chi_n(\textbf{x})]
\end{eqnarray}

Then, by using standard properties of the Fourier transform, Eq.
(\ref{eq:coeff2}) can be formally written in coordinate space as
\begin{eqnarray}
&&
i\hbar\partial_t\chi_n(\textbf{x})=E_n(-i\hbar\bm{\nabla})\chi_n(\textbf{x})
+U(\textbf{x})\chi_n(\textbf{x})\nonumber\\&&
+\frac{\hbar}{m}\sum_{n'\neq n}\sqrt{\frac{\Omega}{(2\pi)^3}}
\int_k{\rm e}^{i\textbf{k}\cdot\textbf{x}}\int_{k'}
\frac{\textbf{P}_{nn'}(\textbf{k},\textbf{k}')}{\Delta
E_{nn'}(\textbf{k},\textbf{k}')} \!\cdot\!\nonumber\\ &&\quad\cdot
(\textbf{k}\!-\!\textbf{k}')
\tilde{U}(\textbf{k}\!-\!\textbf{k}')\varphi_{n'}(\textbf{k}')
\label{eq:chi}
\end{eqnarray}
This equation is equivalent to the generalized form of the Wannier
equations of Ref. \cite{adams}, with the advantage of having the
interband term written in a more transparent form in terms of its
Fourier components.  This expression allows for a simple
manipulation of the above equation, that for practical use, has to
be further simplified.  The simplest approach is to adopt the
following standard approximations \cite{davies}, assuming that
\begin{itemize}
\item[i)] the energy spectrum is of simple form with minima/maxima
of each band at some point $\bm{k}=\bm{k}_0$ in the first
Brillouin zone;

\item[ii)] the $\varphi_n(\textbf{k})$ functions are localized on
a small region of $\bm{k}$ space around $\bm{k}=\bm{k}_0$ during
the whole evolution of the system.
\end{itemize}
For convenience in the notations, and without loss of generality,
in the rest of the paper we will set $\bm{k}_0=0$. Then, let us
consider the term
${\textbf{P}_{nn'}(\textbf{k},\textbf{k}')}/{\Delta
E_{nn'}(\textbf{k},\textbf{k}')}$ that characterizes the interband
coupling; to first order in $k$, $k'$ we can write
\begin{eqnarray}
\label{eq:expk}
&&\frac{\textbf{P}_{nn'}(\textbf{k},\textbf{k}')}{\Delta
E_{nn'}(\textbf{k},\textbf{k}')}= \frac{\textbf{P}_{nn'}}{\Delta
E_{nn'}}\\ &&\quad+ \frac{1}{\Delta
E_{nn'}}\left(\textbf{k}\!\cdot\!{\bm\nabla}_k\textbf{P}_{nn'}
+\textbf{k}'\!\cdot\!{\bm\nabla}_{k'}\textbf{P}_{nn'}\right)+O(k^2)
\nonumber
\end{eqnarray}
where to simplify the notation we have defined $\Delta
E_{nn'}=\Delta
E_{nn'}(\textbf{0},\textbf{0})=E_{n}(\textbf{0})-E_{n'}(\textbf{0})\equiv
E_{n}\!-\!E_{n'}$,
$\textbf{P}_{nn'}=\textbf{P}_{nn'}(\textbf{0},\textbf{0})$, and we
have used the fact that the energies do not contain first order
terms in $\textbf{k}$ (see (ii)).

The first derivatives in Eq. (\ref{eq:expk}) can be evaluated by
using the relation
\begin{equation}\label{svil}
{\bm\nabla_k} u_n(\textbf{k},\textbf{x})= \frac{\hbar}{m}
\sum_{n'\neq n} u_{n'}(\textbf{k},\textbf{x})
\frac{\textbf{P}_{n'n}(\textbf{k},\textbf{k})} {E_n(\textbf{k})-
E_{n'}(\textbf{k})}
\end{equation}
that can be obtained by differentiation Eq. (\ref{eq:eigenu}) with
 respect to $\textbf{k}$, and projecting the term ${\bm\nabla}_k
 u_n(\textbf{k},\textbf{x})$ on the $u_{n}(\textbf{k},\textbf{x})$
 basis \cite{adams}.  Then a straightforward calculation yields
\begin{eqnarray}
{\bm\nabla}_k\textbf{P}_{nn'}&=& \frac{ \hbar}{m} \sum_{n''\neq n}
\frac{ \textbf{P}_{nn''}\textbf{P}_{n''n'} }{E_n- E_{n''}}\equiv
M_{nn'}\\ {\bm\nabla}_{k'}\textbf{P}_{nn'}&=& \frac{ \hbar}{m}
\sum_{n''\neq n'} \frac{\textbf{P}_{nn''}
\textbf{P}_{n''n'}}{E_{n'}\!-\! E_{n''}}= M^*_{n'n}
\end{eqnarray}
where ${\bf P}^*_{n'n}(\textbf{k},\textbf{k}')= {\bf
P}_{nn'}(\textbf{k},\textbf{k}')$.  The last term of Eq.
(\ref{eq:chi}) then becomes
\begin{eqnarray}\label{eq:svil}
&&\int_{k'}\frac{\textbf{P}_{nn'}(\textbf{k},\textbf{k}')}{\Delta
E_{nn'}(\textbf{k},\textbf{k}')} \!\cdot\!
(\textbf{k}\!-\!\textbf{k}')
\tilde{U}(\textbf{k}\!-\!\textbf{k}')\varphi_{n'}(\textbf{k}') =\\
&&\frac{\hbar}{m} \sum_{n'\neq n}\frac{ \textbf{P}_{nn'}}{\Delta
E_{nn'}} \int_{k'} (\textbf{k}\!-\!\textbf{k}')
\tilde{U}(\textbf{k}\!-\!\textbf{k}')\varphi_{n'}(\textbf{k}') \nonumber\\
&& +\frac{\hbar}{m}\sum_{n'\neq n} \frac{ M^*_{n'n}}{\Delta
E_{nn'}} \int_{k'} (\textbf{k}\!-\!\textbf{k}')
\tilde{U}(\textbf{k}\!-\!\textbf{k}') \textbf{k}'
\varphi_{n'}(\textbf{k}') \nonumber\\ && + \textbf{k}
\frac{\hbar}{m}\sum_{n'\neq n} \frac{M_{nn'} }{\Delta E_{nn'}}
\int_{k'} (\textbf{k}\!-\!\textbf{k}')
\tilde{U}(\textbf{k}\!-\!\textbf{k}') \varphi_{n'}(\textbf{k}') +
o(\textbf{k}^2) \nonumber
\end{eqnarray}
This expression allows us to write Eq. (\ref{eq:chi}) as
\begin{widetext}
\begin{eqnarray}
\label{eq:final}
 &&i\hbar\partial_t\chi_n(\textbf{x})=E_n(- i\hbar\bm{\nabla}
)\chi_n(\textbf{x})+ U(\textbf{x}) \chi_n(\textbf{x})
-i\bm{\nabla} U(\textbf{x}) \frac{\hbar}{m}\sum_{n'\neq n}\frac{
\textbf{P}_{nn'}}{\Delta E_{nn'}} \chi_{n'}(\textbf{x})\\ &&\quad
-\bm{\nabla} U(\textbf{x}) \frac{\hbar}{m}\sum_{n'\neq n}
\frac{M^*_{n'n} }{\Delta E_{nn'}} \bm{\nabla}\chi_{n'}(\textbf{x})
- \frac{\hbar}{m} \sum_{n'\neq n}\frac{M_{nn'} }{\Delta
E_{nn'}}\left[\nabla^2 U(\textbf{x})\chi_{n'}(\textbf{x}) +
\bm{\nabla} U(\textbf{x})\bm{\nabla} \chi_{n'}(\textbf{x}) \right]
\nonumber
\end{eqnarray}
\end{widetext}

The above equation represents the main result of this paper.  It
describes the evolution of the Wannier envelope functions by fully
including the effects of the periodic potential and accounting for
the interband coupling due to the perturbation potential $U$ up to
second order in $k$.  Eq. (\ref{eq:final}) can be further
simplified by means of the usual effective mass approximation that
amounts to retaining only up to quadratic terms in $k$ in the
kinetic operator. In general this corresponds to replace the bare
mass by a $3\times3$ mass tensor $m^*_{ij}$ \cite{ashcroft}; in
the special case of an isotropic periodic potential or for a
one-dimensional system as we will consider later, one simply gets
\begin{equation}
\label{eq:mstar} E_n(\textbf{k})= E_n+\frac{\hbar^2
k^2}{2m_n^*}+O(k^3).
\end{equation}

\section{Comparison with other ${\bf k} \cdot {\bf P}$ models}
\label{sec:comparison}

Let us now discuss the connection of the model in Eq.
(\ref{eq:final}) with other two ``${\bf k} \cdot {\bf P}$''
models, namely the Kane and Luttinger-Kohn (LK) models
\cite{kane,luttinger}, which are widely used both to estimate the
band diagram in semiconductor and the transmission coefficients of
interband devices \cite{Chakraborty,Edwards}.

The Kane model is based on the following choice of the basis
elements
\begin{equation}
\langle\textbf{x}|n,\textbf{k}\rangle_{Ka}\equiv {\rm
e}^{i\textbf{k}\cdot\textbf{x}}u_n(\textbf{0},\textbf{x})\;,
\end{equation}
that form a complete orthonormal set, and can be used to expand
the electron wave function $\Psi$ in a similar way of what shown
in the previous section. On this basis the Schr\"odinger equation
takes the form
\begin{equation}
i\hbar\partial_t\varphi_n(\textbf{k})=
\sum_{n'}\int_{k'}\mathcal{H}^{Ka}_{nn'}(\textbf{k},\textbf{k}')\:
\varphi_{n'}(\textbf{k}')
\end{equation}
where the hamiltonian matrix elements are
\begin{eqnarray}
\label{eq:matrix_el_Ka}
\lefteqn{\mathcal{H}^{Ka}_{nn'}(\textbf{k},\textbf{k}')\equiv\langle
n,\textbf{k}|H_0+U|n',\textbf{k}'\rangle_{Ka}}
\\\nonumber&=&
\left[\left( E_n + \frac{\hbar^2 k^2}{2 m_0} \right)
 \delta_{nn'}+ \frac{\hbar}{m_0}
\textbf{k}\!\cdot\! \textbf{P}_{nn'}\right]
\delta(\textbf{k}\!-\!\textbf{k}')
\\\nonumber&&\qquad\;
+\tilde{U} (\textbf{k}\!-\!\textbf{k}')\delta_{nn'}\;.
\end{eqnarray}

By means of an inverse Fourier transform (see Eq.
(\ref{eq:def_env_f})) it is straightforward to recover the
equation for Kane envelope functions $\chi_n^{Ka}(\textbf{x})$
\begin{eqnarray}
\label{eq:kane}
 i\hbar\partial_t\chi_n^{Ka}(\textbf{x})&=&
\left(-\frac{\hbar^2}{2m}\bm{\nabla}^2 + E_n +
U(\textbf{x})\right)
\chi_n^{Ka}(\textbf{x})\nonumber\\
&& -i\frac{\hbar}{m}\sum_{n'\neq n}
\textbf{P}_{nn'}\!\cdot\!\bm{\nabla}\chi_{n'}^{Ka}(\textbf{x})\,.
\end{eqnarray}
This equation shows that in the Kane representation envelope
functions related to different ``band" indices are coupled even if
the external field is vanishing.  This is due to the fact that the
unperturbed hamiltonian $H_0$ is not diagonal on the Kane basis
(see Eq. (\ref{eq:matrix_el_Ka})), and therefore the $n$ here does
not correspond to the usual band index of the Bloch picture. In
other words this means that the envelope functions
$\chi_{n}^{Ka}(\textbf{x})$ do not have the direct physical
meaning of wavefunctions of an electron in a definite energy band.
As a consequence, one should be careful in estimating truncation
errors when the full problem is reduced to a finite set of
envelope functions.

To overcome the previous difficulty, Luttinger and Kohn proposed a
different choice of the basis functions \cite{luttinger}. The idea
is to use a quasi-unitary transformation $\Theta$ to diagonalize
the Kane hamiltonian in the momentum space up to first order in
$k$. In this way, it is possible to get a natural extension of the
effective mass single band model in the multiband framework. The
new hamiltonian reads
\begin{equation}\label{eq:rot_mat_Ka}
\mathcal{H}^{LK}= \Theta ^{-1}  \mathcal{H}^{Ka} \Theta
\end{equation}
where $\Theta$ is defined as follows
\begin{equation}
\langle n,\textbf{k}| \Theta |n',\textbf{k}'\rangle_{Ka} = \left(
\delta_{nn'} -  \frac{\hbar}{m_0} \frac{ \textbf{P}_{nn'}\!\cdot\!
\textbf{k}}{\Delta E_{nn'} }
\right)\delta(\textbf{k}\!-\!\textbf{k}')\,,
\end{equation}
providing a unitary transformation to first order in $k$.
Accordingly, the elements of the LK basis are defined by
$|n,\textbf{k}\rangle_{LK}=\Theta |n,\textbf{k}\rangle_{Ka}$, and
correspond to an expansion of the $u_n(\textbf{k},\textbf{x})$
functions to first order in $k$
\begin{eqnarray}
\label{eq:lkbasis} \langle
\textbf{x}|n,\textbf{k}\rangle_{LK}={\rm
e}^{i\textbf{k}\cdot\textbf{x}} \left[u_n(\textbf{0},\textbf{x})
+\textbf{k} \frac{\partial u_n(\textbf{0},\textbf{x})}{\partial
\bf k}\bigg |_{\textbf{0}}\right]\,.
\end{eqnarray}

In the coordinate space the LK model reads
\begin{widetext}
\begin{eqnarray}
\label{eq:chiLK} i\hbar\partial_t\chi_n^{LK}(\textbf{x})&=&
\left[E_n-\frac{\hbar^2}{2m_n^*}\nabla^2+U(\textbf{x})\right]
\chi_n^{LK}(\textbf{x}) -i\bm{\nabla} U(\textbf{x})
\frac{\hbar}{m}\sum_{n'\neq n}\frac{ \textbf{P}_{nn'}}{\Delta
E_{nn'}} \chi_{n'}^{LK}(\textbf{x}) \\\nonumber &&+
 \frac{\hbar}{m_0} \sum_{n''n'\neq n} \left(
 \textbf{P}_{nn''}\bm{\nabla}\right)
\left(\textbf{P}_{n''n'}\bm{\nabla}\right)
 \left(\frac{1}{\Delta E_{nn''}}\!-\!\frac{1}{\Delta E_{n''n'}}
  \right) \chi_{n'}^{LK}(\textbf{x})
\end{eqnarray}
\end{widetext}
The first line here corresponds to the first line of Eq.
(\ref{eq:final}) with the effective mass approximation
(\ref{eq:mstar}); the second line instead represent a spurious
coupling between different bands that corresponds to the choice of
the truncated basis in Eq. (\ref{eq:lkbasis}), and is usually
neglected \cite{callaway}. In our approach this term would come
from the expansion of the off-diagonal kernel
\begin{equation}
\langle n,\textbf{k}|H_0|n',\textbf{k}'\rangle =\int_{x} {\rm
e}^{-i\textbf{k}\cdot\textbf{x}} u_n^*(\textbf{k},\textbf{x})\:
H_0 \:{\rm
e}^{i\textbf{k}\cdot\textbf{x}}u_{n'}(\textbf{k}',\textbf{x})
\end{equation}
but is cancelled exactly by a term coming from the expansion of
the $u_n(\textbf{k},\textbf{x})$ functions to second order in $k$.
As a matter of fact, this contribution is absent in our approach
since only the off-diagonal terms that depend on the external
potential $U$ have been approximated (up to $O(k^2)$ in Eq.
(\ref{eq:final})).

We also remark that in the LK approach one usually neglect also
the interband coupling proportional to the applied field $\nabla
U$, and this prevents any description of interband tunneling
effects.

\section{An example}
\label{sec:example}

As an application of the model discussed in Sec. \ref{sec:model}
we consider a one-dimensional semiconductor device consisting of a
multilayer heterostructure where only two bands play a relevant
role, namely the ``conduction'' and ``valence'' bands. As a
further approximation we keep the interband terms only to first
order in $k$, neglecting the terms proportional to the matrix
$M_{nn'}$, and adopt the effective mass approximation
(\ref{eq:mstar}). Thus the system in Eq. (\ref{eq:final}) can be
reduced to the following set of coupled equations
\begin{widetext}
\begin{equation}\label{eq:twoband1}\left\{\begin{array}{c}
\displaystyle i\hbar\partial_t\chi_c( {x}) = E_c\chi_c( {x})
-\frac{\hbar^2}{2m_c^*}\nabla^2\chi_c( {x}) +U( {x})\chi_c( {x})
-i{\nabla} U( {x}) \frac{\hbar  {P}}{m E_g} \chi_{v}( {x})
\\ \\
\displaystyle i\hbar\partial_t\chi_v( {x}) = E_v\chi_v( {x})
+\frac{\hbar^2}{2|m_v^*|}\nabla^2\chi_v( {x}) +U( {x})\chi_v(
{x})-i{\nabla} U( {x}) \frac{\hbar  {P}}{m E_g} \chi_{c}( {x})
\end{array}\right.
\end{equation}
\end{widetext}
that depend on four phenomenological parameters: the interband
momentum matrix $P\equiv P_{c,v}=P^*_{v,c}$ (see
\cite{swexu,wenckebach} for a numerical estimate), the energy gap
$E_g\equiv E_c-E_v$, and the effective masses $m_{c,v}^*$ for the
conduction and valence bands respectively ($m_{v}^*=-|m_{v}^*|$).

The total potential can be written as $U(x)=U_h(x)+U_e(x)$, where
$U_e(x)$ is the electrostatic potential generated by the charge
distribution in the device, and $U_h(x)$ accounts for the spatial
dependence of the band edges in the heterostructure. Indeed, in
real heterostructures the conduction and valence band edges depend
on $x$ and to account for this we adopt the point of view of
considering the Bloch spectrum as constant among the layers,
treating the actual spatial dependence as an external potential
applied to the heterostructure bulk. We remark also that here we
are considering only coherent transport, neglecting any
dissipative phenomena like electron-phonon scattering that are not
expected to affect significantly the tunneling process.

The electrostatic potential can be calculated self-consistently by
using the Poisson equation
\begin{equation}
\epsilon\nabla^2U_e(x)=q\rho(x)=q^2[C(x)-n(x)] \label{eq:poisson}
\end{equation}
where the total charge distribution $\rho(x)$ is the sum of the
charge concentration $qC(x)$ of the doping ions plus the charge
distribution $-qn(x)$ of free electrons.

Let us now consider an heterostructure device in contact with a
source and a drain reservoir at a temperature $T$. In the presence
of an incident electron beam with momentum $q$ and energy $E(q)$
injected from the reservoirs into the device, the steady state of
the system is obtained from the solution of the stationary
equations
\begin{widetext}
\begin{equation}
\left\{\begin{array}{cccc} E(q) \chi_c^q(x)&=&E_c\chi_c^q(x)
-\displaystyle\frac{\hbar^2}{2m_c^*}\nabla^2\chi_c^q(x)
+U(x)\chi_c^q(x) -i \nabla U(x) \frac{\hbar P}{m E_g}
\chi_{v}^q(x)
\\ \\
 E(q)\chi_v^q(x)&=&E_v\chi_v^q(x)
+\displaystyle\frac{\hbar^2}{2|m_v^*|}\nabla^2\chi_v^q(x)
+U(x)\chi_v^q(x)-i \nabla U(x) \frac{\hbar P}{m E_g} \chi_{c}^q(x)
\label{eq:stationary}
\end{array}\right.
\end{equation}
\end{widetext}
combined with the Poisson equation (\ref{eq:poisson}). The
equations are solved by approximating the spatial derivative by a
Runge-Kutta method and using a Gummel predictor scheme to reach
convergence \cite{numerics}.

The electronic density $n=n_c+n_v$ is constructed in terms of the
pure state solutions $\chi_c^q$ and $\chi_v^q$ of the above
equations, weighted by the momentum distribution of the incident
beams
\begin{equation}
n(x)=\int_0^{\infty} \!\!\!\!\textrm{d} q~ f_0(q)
\left[|\chi_c^q(x)|^2+ |\chi_v^q(x)|^2\right]
\end{equation}
where $f_0(q)$ is the Fermi-Dirac distribution integrated on the
transverse coordinates \cite{Frens1}.

Similarly the electronic current $J=J_c+J_v$ is calculated as
\begin{equation}
J(x)=\sum_{i=c,v}\frac{\hbar}{2 m_i} \int_0^{\infty}
\!\!\!\!\textrm{d} q~ f_0(q)\textrm{Im}\left[
 \chi_i^q(x)\nabla  \chi_i^q(x)\right]\,.
\end{equation}

To model the charge injected in the device from the source and
drain reservoirs we use transparent boundary conditions. For
example, at $x=0$, in case of electron beam incident in the
conduction band with positive momentum $q$, we have
\begin{equation}
\frac{d}{dx}{\chi_c^q \choose \chi_v^q}
\bigg|_{x=0}=\left(\begin{array}{cc} - i k_i & 0 \\
0 & - i k_{r} \\
\end{array}\right){\chi_c^q \choose \chi_v^q}
+{2 i k_i \choose 0}
\end{equation}
were
\begin{eqnarray}
k_i&=& \sqrt{ \frac{2  m_c^*}{\hbar^2 } [E(q)-E_c]}
\\
k_{r}&=& -i \sqrt{\displaystyle \frac{2 |m_v^*|}{\hbar^2  }
[E(q)-E_v]  }\: \: ;
\end{eqnarray}
the other cases are treated in similar way.

\begin{figure}
\centerline{\includegraphics[width=8cm,clip=]{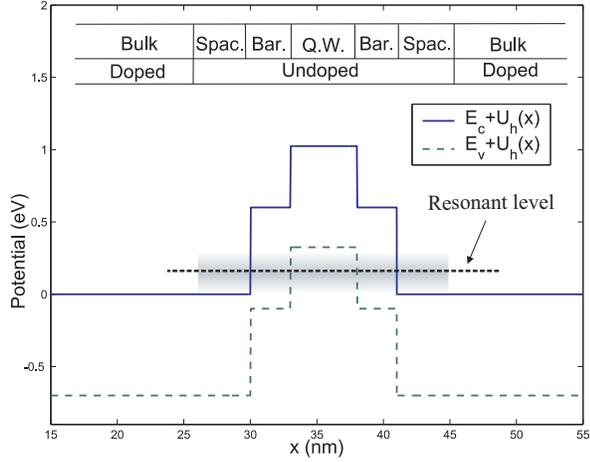}}
\caption{(Color online) Simulated heterostructure profile and doped regions of
the RITD. The widths of the layers are chosen: 5.00 nm for the
quantum well (Q.W.), 3.00 nm for the barriers (Bar.) and 4.50 nm
for the spacer layers (Spac.). Resonant tunneling takes place when
the energy of electrons in the conduction band is resonant with
that of the bounded hole state in the central quantum well (shaded
area).} \label{fig:diag_bande}
\end{figure}

Let us now discuss what happens in case of that is a
one-dimensional RITD, where charge transport across the device
takes place thanks to resonant tunneling between electron states
in the conduction band and hole states in the valence band. As a
specific case here we consider the simple test device depicted in
Fig.~\ref{fig:diag_bande}, consisting of a 5.00-nm-wide quantum
well (for hole states), bounded by two identical 3.00-nm-wide
barriers that are interfaced to the bulk by two 4.50-nm-wide
spacer layers. In Fig.~\ref{fig:diag_bande} we also show the
resulting heterostructure potential in case of a GaSb lattice bulk
with a doping concentration of $10^{18}\: \textrm{cm}^{-3}$ \cite{Kefi}. 
For consistency with the formalism presented in the
previous sections, we are considering a constant band gaps
heterostructure (we remind that in our model all the interband
effects are taken in account by an external potential added to the
bulk Hamiltonian). We note that, as far as resonant tunneling is
concerned, the actual shape of the conduction band far above the
resonance level cannot produce strong effects on the tunneling
process. Therefore we expect that the I-V behaviour does not
change qualitatively in case of more complex conduction bands
profiles.

\begin{figure}
\centerline{\includegraphics[width=8cm,clip=]{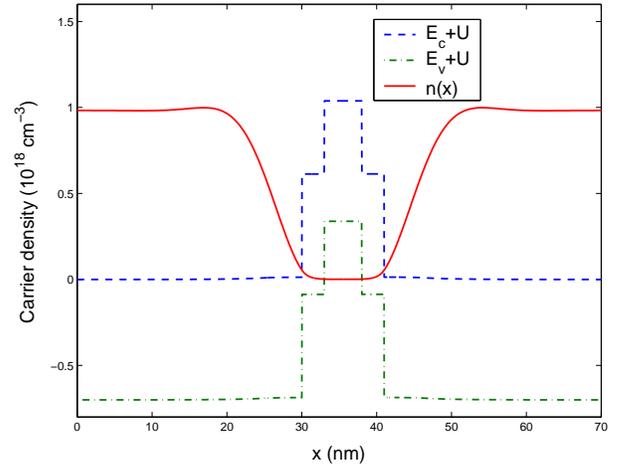}}
\caption{(Color online) Self-consistent potential profile and density of
electrons corresponding to unbiased case.}
\label{fig:densita_equilibrio}
\end{figure}

In Fig.~\ref{fig:densita_equilibrio} we show the calculated
equilibrium self-consistent potential $U$ and the density of
electrons corresponding to the unbiased case: note that, in this
case, the profile of the heterostructure potential $U_h$ of
Fig.~\ref{fig:diag_bande} is practically unchanged by the addition
of the electrostatic potential~$U_e$.

\begin{figure}
\centerline{\includegraphics[width=8cm,clip=]{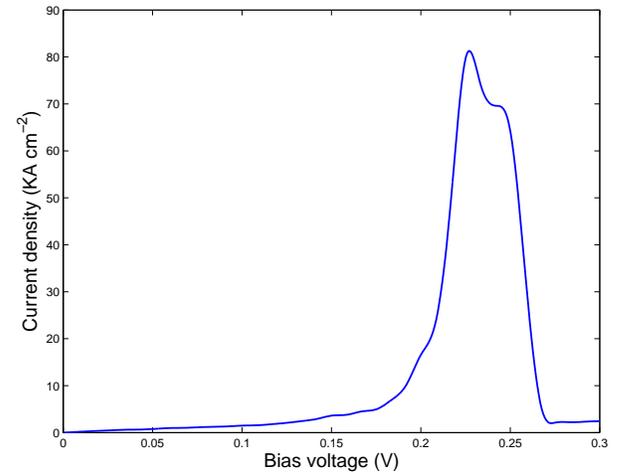}}
\caption{(Color online) $I-V$ characteristic of the simulated diode. Notice the
negative differential resistance for $V_{b}> V_0=0.05$ V.}
\label{fig:diag_I-V}
\end{figure}

The steady I-V characteristic of the device at a temperature of
$300$ $^\circ$K is shown in Fig. \ref{fig:diag_I-V}, where the
current $I$ flowing through the device is plotted as a function of
the bias voltage $V_{b}$ applied to the drain contact. This
picture shows that the model is capable to reproduce the expected
negative differential resistance (NDR) in a certain range of
values of the applied potential (here for $V_{b}> V_0=0.225$ V).

\begin{figure}
\centerline{\includegraphics[width=8cm,clip=]{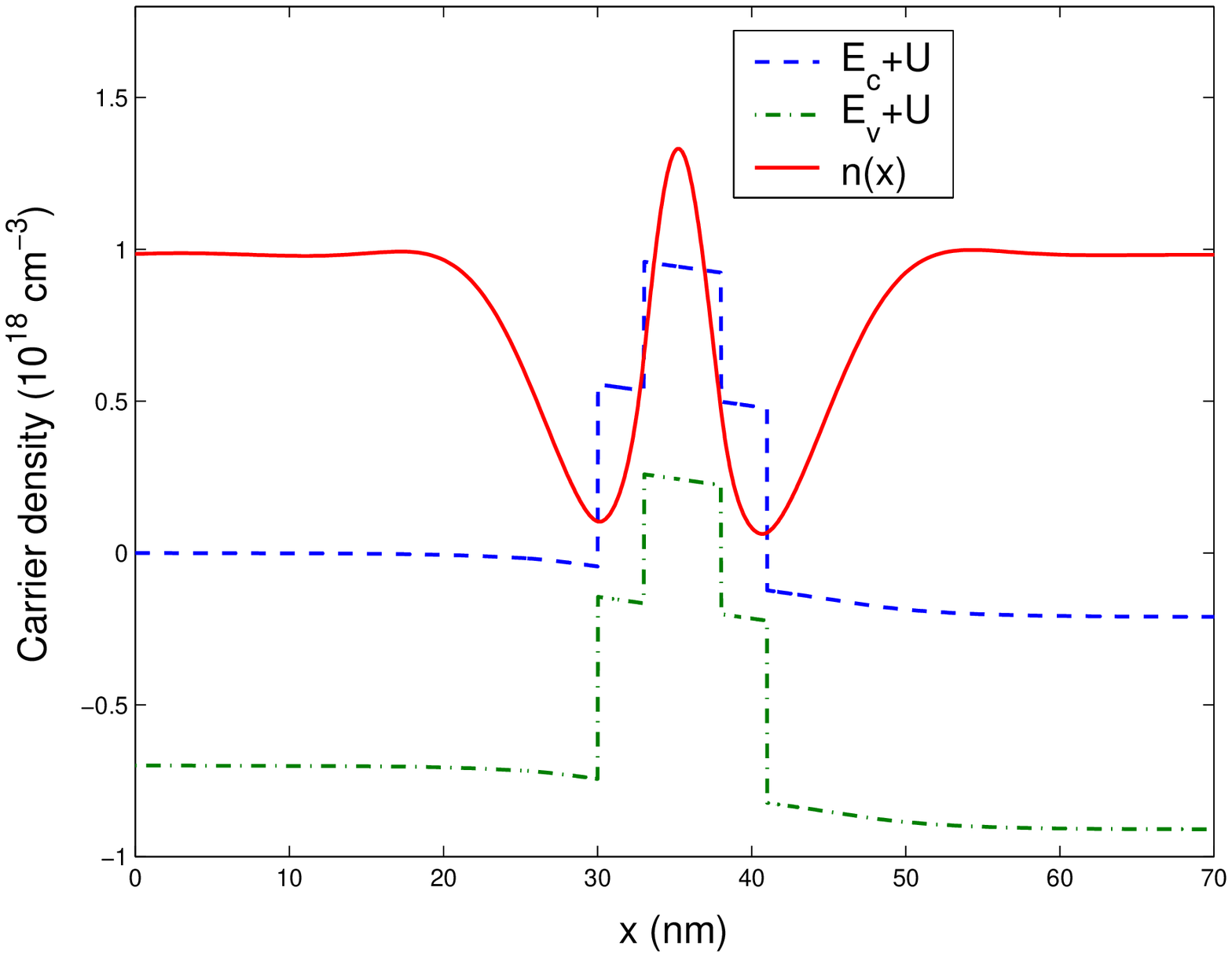}}
\centerline{\includegraphics[width=8cm,clip=]{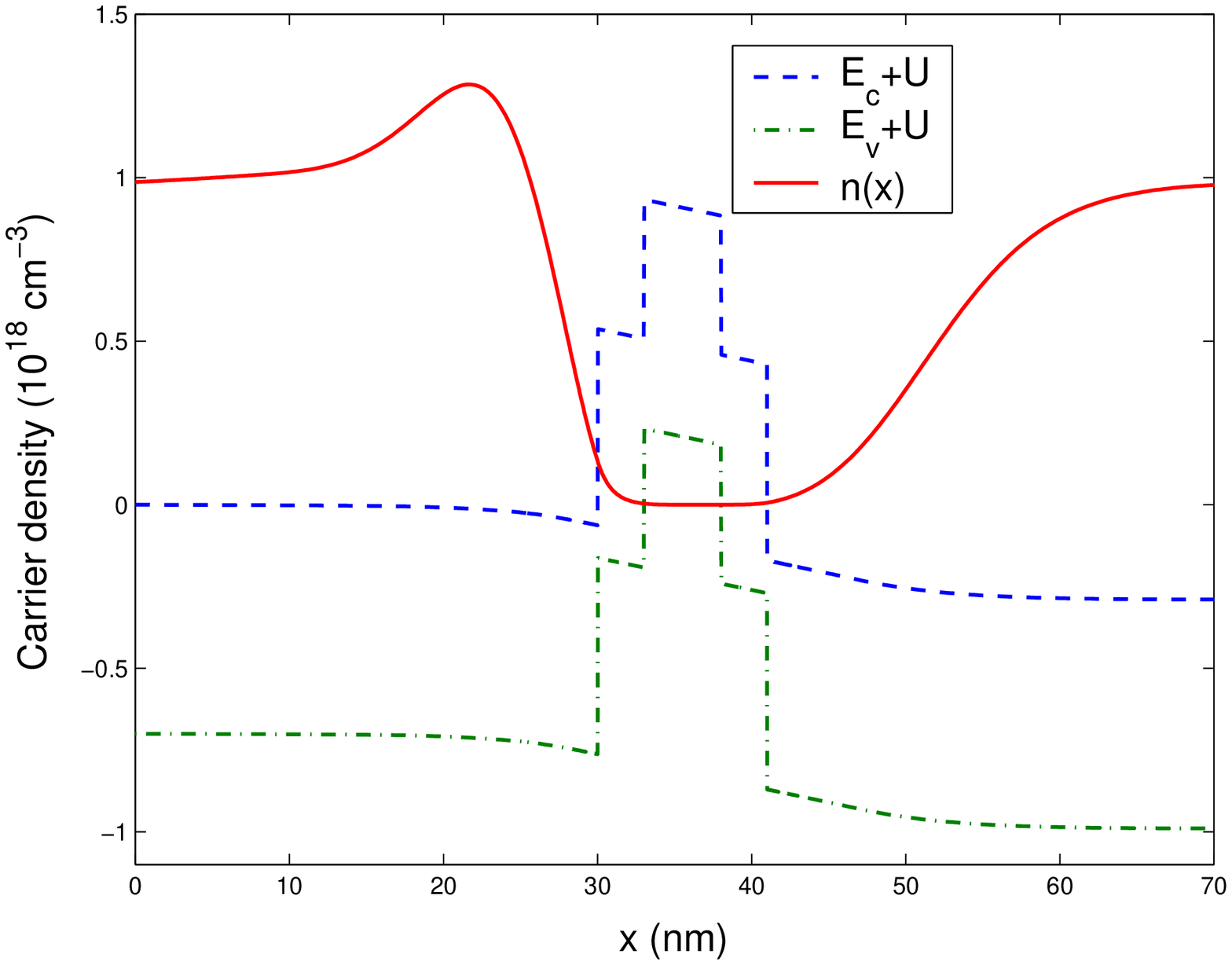}}
\caption{(Color online) Self-consistent potential profile and density of
electrons corresponding to the peak (top) and to the valley
(bottom) currents.} \label{fig:densita_picco_valle}
\end{figure}
In Fig. \ref{fig:densita_picco_valle} we show the calculated
self-consistent potential profile $U$ and the density of electrons
corresponding to peak and valley currents, for $V_{b}=0.225$ V and
$V_{b}=0.27$ V respectively. We note that as expected the density
of electrons in the central region of the potential profile (see
Fig. \ref{fig:densita_picco_valle}) is an increasing function of
the applied bias below resonance ($V_{b}<V_0$), and then sharply
decreases in the NDR region.

These results demonstrate that the present model is promising for
reproducing the I-V curves of physical devices as that considered
e.g. in \cite{Marquardt,Liu}, although a precise comparison
requires the extension to structures with varying band gap
profiles (or, more generally, with $x$-dependent Luttinger-Kohn
parameters \cite{Chao_Chuang}) and will be considered in a future
publication.

\section{Conclusions}
\label{sec:conclusion}

We have presented a multiband model for electron transport in a
crystal lattice. The model is derived within the usual Bloch
theory by means of a $k$-expansion, and is formulated in terms of
cell-averaged envelope functions obtained by projection in the
Wannier representation.  The model is suited to describe in a
clear fashion tunneling effects between different bands in
presence of an applied potential.  Its advantages with respect to
other widely used approaches has been discussed.

As an application we have considered the case of a RITD, a
heterostructure device where the electronic current flows between
a ``conduction" and a ``valence" band, interfaced by potential
barriers. In this case the model is reduced to a system of two
Schr\"odinger equations for the electron envelope function coupled
with the Poisson equation for the field generated by the
electronic distribution itself. It nicely reproduces the expected
behaviour of the current as a function of the applied voltage,
exhibiting a negative differential resistance in a certain range
of values of the applied bias.

The present approach is therefore particularly promising for
reproducing the behavior of physical devices characterized by
resonant tunneling, and we are currently working at the extension
of the model which takes in account the specific symmetry
properties of the crystal lattice and to treat degenerate and 
varying band gap profiles.  Moreover, with
the inclusion of spin-orbit coupling, our model may also be
relevant for the field of spintronics, where ${\bf k} \cdot {\bf
P}$ methods are becoming increasingly important
\cite{spintronics1,spintronics2}.

\begin{acknowledgments}
We thank G. Frosali, G. Borgioli, J. Kefi and N.~Ben~Abdallah for
many stimulating discussions. O. M. acknowledges hospitality from
the MIP laboratory of the P. Sabatier University of Toulouse.
\end{acknowledgments}

\end{document}